# CRYPTAXFORENSIC, WHEN CRYPTOCURRENCY, TAXATION, AND DIGITAL FORENSIC COLLIDE: AN OVERVIEW OF INDONESIAN CRYPTOCURRENCY MARKET

Dimaz Ankaa Wijaya[a], Dony Ariadi Suwarsono[b]

[a] Monash University, Australia Email: dimaz.wijaya@monash.edu
[b] Directorate General of Taxes, Indonesia Email: dony.suwarsono@pajak.go.id

**ABSTRACT**

*Blockchain has emerged into one of the most promising technologies for the future. Its most successful implementation in the form of cryptocurrency has shifted many existing paradigms where financial instruments were limited by locations or jurisdictions. While blockchain is touted to offer many significant and promising features on the other hand it also increases the difficulty level in the taxation area as well as digital forensics. We investigated the issues and explores the real-world situation and how taxation and digital forensics can cope with these technology challenges.*

*Blockchain telah berkembang menjadi salah satu teknologi yang paling menjanjikan di masa mendatang. Implementasi blockchain yang paling sukses saat ini, mata uang kripto, telah mulai memberikan dampak perubahan paradigma yang ada, di mana instrument keuangan tradisional terbatas pada waktu dan yurisdiksi. Blockchain digadang-gadang dapat menawarkan fitur-fitur yang signifikan dan menjanjikan, namun di sisi lain dapat meningkatkan kesulitan dalam bidang perpajakan dan juga forensik digital. Kami menginvestigasi persoalan ini dan mengeksplorasi kondisi nyata yang ada, termasuk pula mempelajari bagaimana perpajakan dan forensic digital dapat beradaptasi terhadap tantangan teknologi ini.*

## 1. INTRODUCTION

The launch of Bitcoin in early 2009 marks the birth of the cryptocurrency industry. The field has grown rapidly in the last few years that now its market value reaches several hundred billion US dollar[1]. Bitcoin as a brand has been known globally as it has been in the spotlight of mainstream media platforms as well as other media managed personally. Its staggering price increase motivates people to trade Bitcoin for fiat currencies such that new cryptocurrency exchanges are now a profitable business.

Although it is now extremely popular, the original writer of Bitcoin's whitepaper as well as its first version of the codes is only known by the name Satoshi Nakamoto, a pseudonym. He improved the existing idea of implementing cryptographic techniques into a payment system without any central party as written in the whitepaper (Nakamoto 2008). He also coined the term Bitcoin as well as the first to "mine" Bitcoin when creating the first block referred as genesis block.

The success of Bitcoin has sparked a massive development in the cryptocurrency area. Currently, there are more than 1,600 cryptocurrencies

---

[1] Total Market capitalisation of US$212B according to Coinmarketcap on 6 November 2018

available in the market[2], each with its uniqueness, strengths, and weaknesses. This new technology is also expected to affect financial industry as well as taxation, where the digital asset can be utilized not only to store value but also to send payments.

**Our Contribution**. First, we start with explaining the technology and develop classifications of cryptocurrencies based on their characteristics. Then we explore cryptocurrency usages and relate the matter in taxation area. We also explore how digital forensics can be done to gather information regarding cryptocurrency activities. Lastly, we suggest several steps to enhance cryptoforensics which are tailored for law enforcement agencies such as Directorate General of Taxes.

## 2. BACKGROUND

This chapter describes the background knowledge of our paper, which consists of three parts, namely cryptocurrency, taxation, and digital forensics.

### 2.1. Cryptocurrency
### 2.1.1. Overview

Instead of relying on a central authority to control and run the whole system, cryptocurrency lets anyone to conduct data verification on a publicly available distributed ledger. To enable the feature, several cryptographic techniques are implemented in the system, making it fully run by the automated protocols.

Cryptocurrency runs on the Internet, where traditional boundaries such as working hours, geographical locations, and jurisdictions are eliminated. International money transfers can be done through cryptocurrency network which can be verified within minutes. The success of Bitcoin has caused a rapid adoption of cryptocurrencies across the world.

### 2.1.2. The Technologies Behind Cryptocurrency

In recent days, there is a huge collection of technologies being implemented in various cryptocurrencies. However, there are several core technologies that are used in (almost) all cryptocurrencies.

**Public Key Cryptography** (PKC). PKC is a cryptographic technique which utilises two keys: public key and private key. As the name implies, the public key is intended to be publicized (shared) to other parties, while the private key needs to be kept secret by the owner. Both keys are different but having a 1-to-1 relationship. This technique enables the cryptocurrency users to prove the ownership of the coins without revealing their real identities by the help of digital signatures.

**Distributed Ledger Technology** (DLT). DLT is a public ledger where the information stored in the ledger is shared among system participants. The information inside DLT is constructed such that stored data is infeasible to tamper. DLT is an append-only database; consensus method is applied to determine which participant can write the next data. There are at least two major types in DLT, namely blockchain and Directed Acyclic Graph (DAG). In DLT, new data is first collected and then stored once every predetermined period (epoch), constructing a sequential historical data. The way it is stored in the existing storage is by linking it to the previous data: the hash of the previous data is included in the new data. While blockchain can only have one main branch in the system (Nakamoto 2008), DAG allows multiple branches when storing information (Popov 2016). In this paper, blockchain and DLT will be used interchangeably to refer DLT as a general concept.

---

[2] Based on information published by Coinmarketcap.com on 6 November 2018

**Hash functions and Merkle Tree**. Hash function is an algorithm to convert an arbitrary length data into a value which has a constant length over the same hash function. One of the main roles of hash function is to represent data without mentioning the data at all, hence it acts as the commitment of the represented data. Merkle Tree is a form of binary tree where anyone can verify the correctness of any data within its leaves without the need of knowing all the leaves. The Merkle Tree is often used as an authenticated data structure (ADS) where it is easy to verify the integrity of its leaves.

### 2.1.3. Various Types of Cryptocurrency

Classifying cryptocurrencies is important in digital forensics as it will determine the tools and methods during the investigation, or even determine how difficult the investigation will be. Based on its characteristics, we can classify the cryptocurrencies into several types. The naming convention is coined based on Bitcoin as the standard cornerstone, hence we also use the blockchain technology in our classification.

**Blockchain 1.0**. The cryptocurrencies in this class are those cryptocurrencies following the identical (or with minor changes) protocols of Bitcoin. Litecoin and Dogecoin are two most popular cryptocurrencies other than Bitcoin. This class is marked by its legacy protocols such as Nakamoto Consensus, pseudo-anonymity, and input-output transactions. The main function of these cryptocurrencies is payment method.

**Blockchain 1.1**. This class is for those cryptocurrencies that have evolved from version 1.0 by embedding additional features on their systems. The cryptocurrencies in this class are NXT, Ardor, NEM, Waves, and others. Not only as payment methods, these cryptocurrencies also provide rich features such as peer-to-peer marketplace, child chains, digital asset management, electronic voting, and so on.

**Blockchain 1.2**. The need of stronger anonymity motivated people to embed extra anonymity features on cryptocurrencies. The members of this class are known to be more difficult to analyse than Bitcoin, such as Monero, Zcash, and Dash. Several other cryptocurrencies such as Verge claimed that these products are anonymous, despite it is still unclear how their proposed protocols can provide a strong anonymity to the data and to their users.

**Blockchain 2.0**. The development of Ethereum has expanded the scope of blockchain from merely storing data into a system that enables its users to run programs on its virtual environment which is in sync to its permanently-stored data. Blockchain 2.0 offers smart contract capability to its users, where the users are given the flexibility of creating complex algorithms using the provided software development tools. Not only providing a payment scheme, the smart contract can be utilised to develop decentralized applications (**dApps**) with unlimited features which will be run directly on the blockchain (Wood 2014).

**Token**. This class exists due to the popularity of smart contract. In a smart contract platform, a developer can create a contract which contains values, or tokens. These tokens can be set to be transferred from their owners to receivers. Token can be treated as a different class of cryptocurrency, but it can also be a derivation of smart contract. The most popular token platform at the time of the writing is Ethereum, where ERC20 (Vogelsteller and Buterin 2015) is the most known standard when creating tokens. By implementing the ERC20 standard, a token can be traded with other tokens in a peer-to-peer token market as they have identical functions and characteristics. This type of token is also commonly called as ERC20 token.

### 2.1.4. Activities Related to Cryptocurrency

As a system, several activities are closely related to cryptocurrency. These activities are: mining/staking, transacting, trading, and providing services. Mining and staking are activities related to cryptocurrency's open consensus. In a cryptocurrency implementing a Proof-of-Work (PoW) consensus, a miner uses computing devices such as computers and Application-Specific Integrated Circuit (ASIC). The miners compete to compute mathematical questions, where the winner can write the next block. Stakers are participant of any cryptocurrencies implementing Proof-of-Stake (PoS) consensus mechanism, where coins are being staked and raffled to determine who has the authority to produce the next block. Both miner and staker are motivated by the prize coin promised to be rewarded by the protocol as well as the right to claim the transaction fees paid by the cryptocurrency users.

While mining requires investment in the form of computing power and staking requires a huge amount of money to acquire a significant portion of the coins, creating transactions can be done by anyone. The two requirements for creating a transaction are a wallet and a fraction of coin. Creating a transaction in a cryptocurrency system means the user utilises the wallet application to construct a digital information which expresses the coin movement from the payer to the payee(s). This information will be authorized by using digital signature(s) which will validate that the payer owns the coin(s) to be sent to the payee(s). It is feasible to send a fraction of a coin only if the amount exceeds the minimum amount (if determined by the cryptocurrency system) and sufficient to pay the transaction fee.

Cryptocurrency trading activity is a logical result of the emerging digital asset. As the cryptocurrencies were getting more valuable and usable in daily life, cryptocurrency trading facilitates supply (ask) and demand (bid). While the trading platform will receive trading fees, the traders expect to make profits from their trading activities.

As a valuable intangible asset, it is possible that the cryptocurrencies are used as a payment method, as most of them were intended to be. However, until now the Indonesian government do not facilitate the use of cryptocurrencies as a legal tender, where Currency Act 2011 defines Rupiah as the only legal tender in Indonesian jurisdiction. The statement is further clarified by Indonesian Central Bank through a press release on 13 January 2018 (Press Release No. 20/4/DKOM) with a lengthy explanation through a FAQ document[3].

### 2.1.5. Cryptocurrency-related Tools

In general, a cryptocurrency system is entirely run by its set of protocols. The protocols are implemented in one or more application (software) bundles, where there will be multiple and separated parties are running the main application, usually called as a node. Each node manages a copy of the blockchain containing the entire transactions from the genesis block to the newest block. It also validates new transactions, (re)publish transactions to its connected peers, and lets wallets request information about the blockchain data which will be responded with the requested information by the node to the requesting wallets.

A wallet is a client-side application which helps a user to create transactions, manage private keys, and compute the remaining coin balance. The wallet does not store any information about the blockchain, hence it relies on the node when it requires certain information. There are multiple types of wallet, namely mobile wallet which runs in smartphones, desktop wallet in computers, web wallet which can be accessed through web browsers and hosted by service providers. There are also special wallets such as hardware wallet which contains secured storage to protect the private keys and paper wallet which stores private keys in a piece of paper.

Most wallets (excluding paper wallet) are SPV-wallets (Simplified Payment Verification wallet) which only store the blockchain header. The correctness of the information received from the node it is connected to will be checked by using the blockchain header it manages. Merkle tree and the chained blocks structures help the data integrity checking. The relationship between nodes and wallets can be expressed in Figure 1.

---

[3] https://www.bi.go.id/id/ruang-media/siaran-pers/Documents/FAQ_Virtual_Currency_150117.pdf

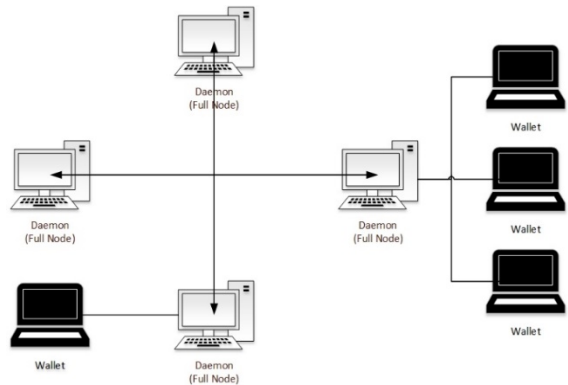

*Figure 1. An illustration of a Blockchain system with multiple nodes (daemons) and several wallets connected to one of the node (Wijaya, et al. 2018).*

A blockchain explorer is an application which will help the users to explore the content of a blockchain. Normally, every cryptocurrency provides at least one blockchain explorer as a way to prove the transparency of the system. However, popular cryptocurrencies such as Bitcoin and Ethereum have many options of blockchain explorers which are provided for free by multiple parties.

### 2.1.6. Indonesian Cryptocurrency Markets

Indodax[4] (previously known as Bitcoin Indonesia) is the first cryptocurrency exchange in Indonesia and the only cryptocurrency exchange for several years. Due to the increasing popularity of cryptocurrencies in Indonesia, now there are more platforms for cryptocurrency trading (with order book), such as Luno[5], Tokocrypto[6], Coinone[7], and Tokenomy[8], including over the counter (OTC) platforms such as Triv[9].

Indodax provides two types of markets, namely IDR (Indonesian Rupiah) and BTC (Bitcoin). In IDR market, the users can buy or sell the paired cryptocurrency by using Rupiah, while in BTC by using Bitcoin. Tokenomy supports four markets (TEN, BTC, ETH, and USDT). Luno pairs Bitcoin with several currencies such as EUR (European Euro), IDR, MYR (Malaysian Ringgit), NGN (Nigerian Naira), ZAR (South African Rand), and ETH (Ethereum). Tokocrypto only sells Bitcoin and Ethereum where both are paired to IDR, while in Coinone Bitcoin, Bitcoin Cash, Litecoin, Ethereum, Ethereum Classic, Qtum, Ripple, and Vexanium are traded with IDR. As an OTC platform, Triv only sells and buys BTC and ETH. Complete lists of cryptocurrencies traded in Indodax and Tokenomy are provided in the Appendix.

Know Your Customer (KYC) and Anti Money Laundering (AML) procedures are commonly implemented in those cryptocurrency exchanges, including those operating in Indonesia. Users are free to create new accounts, but they need to provide real identities (as well as other validation methods such as signatures and self-photos) before the new accounts are activated. The KYC and AML procedures allow the service providers to store the users' identities and ID cards on their own local storage.

---

[4] https://indodax.com
[5] https://www.luno.com/id
[6] https://tokocrypto.com
[7] https://coinone.co.id
[8] https://tokenomy.com
[9] https://triv.co.id

## 2.2. Taxation

Taxation plays an important role in Indonesia State Budget. In financial year 2018, at least 85% of the total revenue for the national budget comes from tax (Kementerian Keuangan Republik Indonesia 2018). This figure shows the importance of taxation for the country. Self-assessment policy has been implemented as a part of the first Indonesian tax reform in 1981 to replace the legacy tax administration system which was highly inefficient and low compliance (Prasetyo n.d.).

However, the self-assessment policy does come with a cost. Tax avoidance cases could happen as the taxpayers tend to minimize their tax due. Tax audits will be established to measure the compliance, and when preliminary proofs are found, the case can be escalated into tax fraud case, which is an offense according to Taxation Act 2008.

## 2.3. Digital Forensics Techniques

Digital forensic has multiple branches according to various medium to store or to compute digital information. Nevertheless, disk-based forensics and network forensics are popular due to their wide scope and usefulness in during the investigation. In the disk-based forensics, the focus of the analyses includes but not limited to operating system analysis, deleted file analysis, password recovery, and user-related data. While the network forensics' focus is on how multiple computers communicate through computer networks, including how network protocols are utilized to carry information from the sender to the receiver.

Digital forensic activities are supported by various forensic tools in the form of hardware and software. In accordance to the activities, there are also protocols and standard procedures to be followed such that the digital information gathered during acquisition process is well-preserved, hence it can be presented as a valid evidence in the court.

## 3. CRYPTAXFORENSIC: CRYPTOFORENSICS FOR TAXATION PURPOSES

In this chapter, we describe cryptaxforensics in details. We coined the term cryptaxforensic to define cryptoforensic activities in the taxation area. Data collection methods, challenges, and proposed methods to enhance cryptoforensics are covered in this chapter.

## 3.1. The (Potential) Role of Cryptoforensics in Directorate General of Taxes

Cryptoforensic techniques have increased in demand and popularity in recent years due to the massive increase of cryptocurrency use. While cryptocurrency was not created for ill-purposes, people always find ways to utilize the medium in legal offences. Tax offense and other financial fraud cases, unfortunately, are not unfamiliar cases in cryptocurrency area, for example in (Pearce 2018). Moreover, cryptocurrencies are considered as tax havens due to their technologies (Marian 2013).

While traditional tax fraud cases will be one of the main concerns for Directorate General of Taxes, cryptocurrency-related cases must also be on the list. Cryptoforensics as digital forensics methods and procedures can be useful in collecting information related to cryptocurrency activities.

## 3.2. Data Collection in Cryptoforensics

We have identified several methods for data collection in cryptoforensics, including disk-based cryptoforensics, network cryptoforensics, blockchain cryptoforensics. Wallet seizure will be discussed in a separate subchapter as it serves a different purpose.

### 3.2.1. Disk-based Cryptoforensics

Disk-based cryptoforensics focuses on the examination of wallet applications which can be used to manage their coins and/or create transactions. As it has been described that there are various types of wallet, multiple digital forensics techniques are also needed to gather information. The desktop wallets are generally compatible with multiple operating systems such as Windows, OSX, and Linux. Moreover, multiple wallets store information in different methods.

However, it is almost certain that the database file to store the private keys and transaction details is encrypted and password-protected. Despite the difficulty to read the private keys, it is advisable to collect all information from the application folders as well as other locations such as home folder and application data folder to investigate the log files. Passwords, PIN keys, or seed words might need to be obtained from the users through any appropriate procedures.

Web browsers such as Mozilla Firefox and Google Chrome are also good places to search. Two most popular Ethereum wallets, MyEtherWallet[10] (MEW) and Metamask[11] provide web browser addons for their clients. In MEW, addresses are visible without login to the wallet, as shown in Figure 2.

---

[10] https://www.myetherwallet.com
[11] https://metamask.io

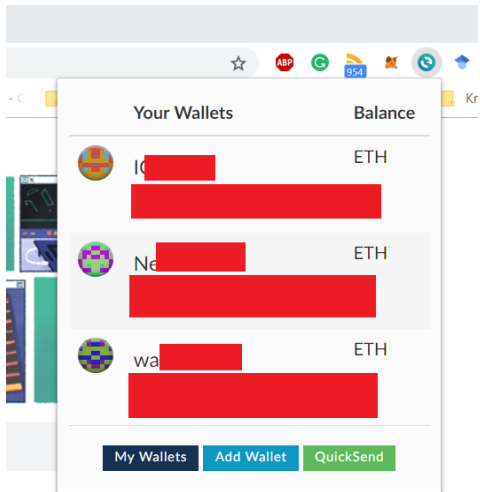

*Figure 2. MyEtherWallet shows several addresses it manages.*

Once the Ethereum addresses are known, the balance will be easily defined by the help of any Ethereum explorer, for example Etherscan[12]. Not only the balance, their associated ERC20 tokens can also be concluded by using the same tool.

### 3.2.2. Network Cryptoforensics

As with other standard network forensic methods, network cryptoforensics also can be done through packet capture techniques. In most cryptocurrency systems, there will be two types of network packets, namely RPC and P2P connections. RPC is the protocol for the connection between a node and a wallet. The wallet sends requests by using RPC interface, and the responses will be sent by the node, either in JSON (JavaScript Object Notation) or BIN (binary) formats.

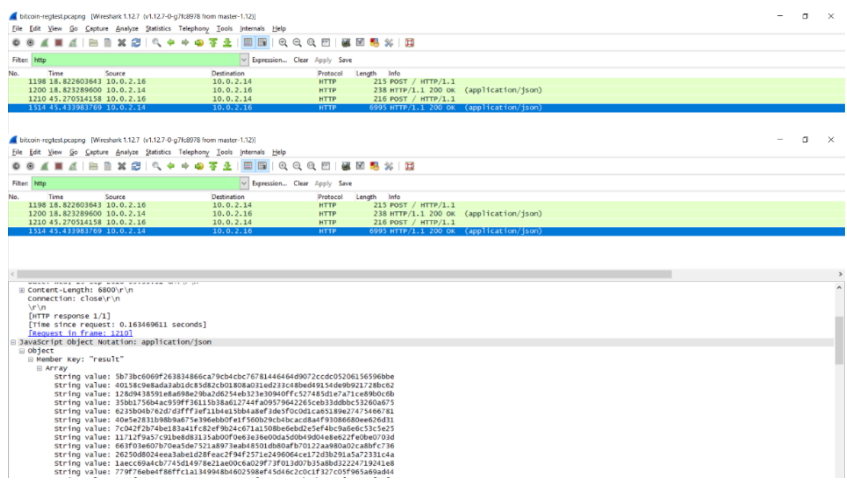

*Figure 3. Captured network packets showing RPC communication.*

---

[12] https://etherscan.io

P2P (peer-to-peer) protocol, on the other hand, is used to exchange information between nodes. The protocol's packets are identified by its "packet magic" as a fingerprint. Every cryptocurrency system has its own packet magic, or often called as "magic bytes". An example is shown on Figure 4.

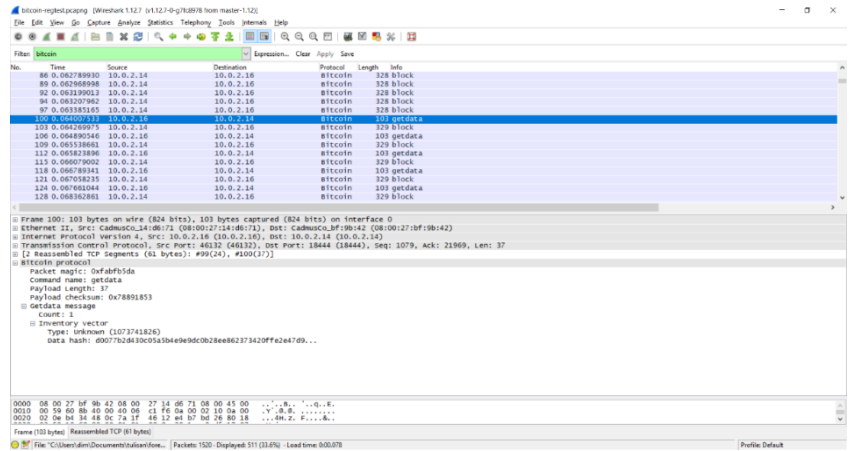

*Figure 4. Captured network packets showing P2P communication.*

### 3.2.3. Blockchain Cryptoforensics

The increase use of cryptocurrency motivated the development of blockchain analyses. Taint analysis is commonly used in Blockchain 1.0, where all related addresses can be concluded. Taint analysis[13] shows other addresses and transactions that are directly and indirectly related to a specific address. However, the taint analysis will eventually produce a cluster which contains a massive number of related addresses, as they are mostly using "haircut" method (Greenberg 2018). Several companies such as Chainanalysis[14], Coinfirm[15], and Ciphertrace[16] use the method to track bitcoins (Greenberg 2018). Furthermore, clustering analyses have been developed and being used to check the information stored in Bitcoin's blockchain (Meiklejohn, et al. 2013) (Ron and Shamir 2013).

A recent research describes a novel approach which was based on a British legal system (Anderson, Shumailov and Ahmed 2018). The new approach describes First-In-First-Out (FIFO) method instead of taint analysis, which results in a shortened list of related address.

### 3.2.4. Wallet/Coin Seizure

Unlike any other digital assets or financial instruments where user IDs are required to transact, cryptocurrencies only require private keys to move coins to other addresses. When the private keys are seized from the users or suspects, it

---

[13] An example of Bitcoin taint analysis can be seen in https://www.walletexplorer.com
[14] https://www.chainanalysis.com
[15] https://www.coinfirm.io
[16] https://ciphertrace.com

is advisable that the information is carefully managed such that anyone having access to the private keys cannot easily move the digital assets. The event where rogue law enforcement agents moved seized bitcoins has happened in the past (Redman 2017)

The way the problem can be mitigated is to create a new wallet with multisignature feature. Several cryptocurrencies such as Bitcoin has this feature, although unfortunately not all cryptocurrencies have implemented it. In a multisignature wallet, each transaction requires more than one valid signature which implies that every transaction is approved by multiple parties. Of course, it is assumed that the new multisignature wallet's private keys are held by different parties.

### 3.3. Challenges

We identify several challenges for cryptoforensics activities. Sometimes, they are not straightforward as its counterparts. Some of these challenges are pseudonymity, privacy-preserving cryptocurrencies, and worldwide markets.

### 3.3.1. Pseudonymity

Although the public blockchains utilized by the cryptocurrencies are transparent and accessible through blockchain explorers, it does not mean that determining the owners of the coins being transacted in the blockchains is trivial. Pseudonymity is a default characteristic in those public blockchains, where the users are identified by their addresses. However, complications occur because one user can create unlimited number of addresses. A quantitative analysis has successfully determined the activities conducted by a cryptocurrency thief (Ron and Shamir 2013), however the same technique cannot be applied to other cases.

### 3.3.2. Privacy-preserving Cryptocurrencies

Privacy-preserving cryptocurrencies such as Monero and Zcash have implemented extra cryptographic techniques such that several information in the blockchain is obfuscated. Monero is using ring signature method where the real signer is obfuscated by using decoys or false addresses. Moreover, the implementation of Ring Confidential Transaction (RingCT) encrypts the amount of coins such that it cannot be seen by anyone but the payer and the payee (Noether and Mackenzie 2016). Several studies have discovered the weaknesses in Monero system (Moeser, et al. 2018) (Kumar, et al. 2017) (Wijaya, et al. 2018), and since then Monero has gone through several updates to improve its security.

Different compared to Monero, Zcash is using zero knowledge proof (ZKP) technology, zk-SNARKS to be specific, where transaction history is completely obfuscated and relationships between coins are hardly identified (Ben-Sasson, et al. 2014). However, recent finding discovered the weakness of the implementation of Zcash (Quesnelle 2018) (Kappos, et al. 2018) due to the extensive computing requirements for its wallet which hinders its users from using the secure mode, namely `z-addr`. Despite this finding, it is difficult to say that Zcash technology has been broken, as the problems found are in the system level.

### 3.3.3. Worldwide Markets

Although there are already exist several cryptocurrency markets in Indonesia, it does not mean that all Indonesian cryptocurrency users are only using them when conducting trading. A trader is motivated to open accounts in as many exchanges as possible, where it is free to create them; all she needs to do is to provide a username and password followed by identity verification.

Cryptocurrency market is unique and totally different compared to stock market, such that there will be multiple prices in different markets. A skilled trader can leverage the opportunity by buying on the lowest price market and selling on the highest price market. By doing this, it is expected that she could gain significant profits.

However, this so-called opportunity for the traders could become a significant obstacle for law enforcement agencies. Unlike any fiat currencies where their transactions are limited by regulations and boundaries, cryptocurrencies do not share the same limitations. A user can send payments through cryptocurrency to the other side of the world in a matter of several minutes. As there are many cryptocurrency markets in the world, a trader can store her digital assets not in the local cryptocurrency exchanges but in any exchanges beyond the reach of any local law enforcement agencies.

## 3.4. Enhancing Cryptoforensics

After identifying several challenges for cryptoforensics, we propose methods to enhance cryptoforensics. Our proposed methods are not about enhancing technical capabilities, but mainly about legal instruments, collaboration, and extensive research.

### 3.4.1. Legal Instruments

We propose the use of any available legal instruments to support cryptoforensics activities. The legal instruments could ensure that information related to cryptocurrency activities within the country's jurisdiction can be acquired. This includes but not limited to enforcing KYC/AML procedures for onboarding new customers of cryptocurrency-related businesses.

### 3.4.2. Collaboration

Collaboration is important in cryptoforensics, especially for gathering a proper information. The main actors for collaboration in this matter are cryptocurrency markets or exchanges. They hold important information about the users, as the users provided their real identities to the markets and this information can be utilized to connect cryptocurrency-related activities with the persons doing them.

An important scheme developed in the Monero community called `EABE` shows the importance of cryptocurrency exchanges (Holmes 2017). `EABE` stands for *Exchange-Alice-Bob-Exchange*. In this scheme, Alice purchases several cryptocurrency coins from an exchange, then she transfers the coins to Bob. Bob then sells the coins to an exchange (not necessarily the same exchange where Alice purchased the coins) to get the local currency. In the scenario, although normally it is infeasible to determine who Alice and Bob are, with the help of two cryptocurrency exchanges, it is now possible to map the transactions and identify two parties, Alice and Bob.

### 3.4.3. Extensive Research

With the rapid development in cryptocurrency technology, digital forensics need to keep up with the pace, hence extensive research in cryptoforensics field need to be conducted. The research focuses on finding relevant artefacts that can help law enforcement agencies to track down illicit activities. Not only in technical area, research can also be conducted on social science, public policies, and economy which will help investigating the correlation between cryptocurrency transactions with real world situations.

## 4. CONCLUSION

We have described cryptaxforensics, a term to express a combination of cryptocurrency, taxation, and digital forensics. The massive adoption of cryptocurrency, the growing number of new technologies that have brought new products into the markets, and the significant rise in the prices could shift the taxation problem from its traditional landscape into a global, broader, and more complex ecosystem as in the cryptocurrency.

We also have described the recent development in both areas: cryptocurrency and digital forensics, including how existing digital forensic methods can be utilized to explore digital evidences with regards to cryptocurrency-related activities. We then discussed challenges in the areas, including proposals to enhance cryptoforensics.

# APPENDIX

*Table 1. The list of cryptocurrencies traded in Indodax. Data taken from Indodax.com on 5 November 2018.*

| No | Cryptocurrency | Code | Market | No | Cryptocurrency | Code | Market |
|---|---|---|---|---|---|---|---|
| 1 | The Abyss | ABYSS | IDR | 17 | Litecoin | LTC | IDR, BTC |
| 2 | Achain | ACT | IDR | 18 | Pundi X | NPXS | IDR |
| 3 | Cardano | ADA | IDR | 19 | NXT | NXT | IDR, BTC |
| 4 | Aurora | AOA | IDR | 20 | Ontology | ONT | IDR |
| 5 | Bitcoin Diamond | BCD | IDR | 21 | SiaCashCoin | SCC | IDR |
| 6 | Bitcoin Cash | BCH | IDR | 22 | Storiqa | STQ | IDR |
| 7 | Bitcoin | BTC | IDR | 23 | Sumokoin | SUMO | IDR, BTC |
| 8 | Bitcoin Gold | BTG | IDR | 24 | Tokenomy | TEN | IDR, BTC |
| 9 | BitShares | BTS | IDR, BTC | 25 | Tron | TRX | IDR |
| 10 | Dash | DASH | IDR, BTC | 26 | Tether | USDT | IDR |
| 11 | Daex | DAX | IDR | 27 | Vexanium | VEX | IDR |
| 12 | Dogecoin | DOGE | IDR, BTC | 28 | Waves | WAVES | IDR |
| 13 | Ethereum Classic | ETC | IDR | 29 | NEM | XEM | IDR, BTC |
| 14 | Ethereum | ETH | IDR, BTC | 30 | Stellar Lumens | XLM | IDR, BTC |
| 15 | Global Social Chain | GSC | IDR | 31 | Ripple | XRP | IDR, BTC |
| 16 | Ignis | IGNIS | IDR | 32 | Zcoin | XZC | IDR |

*Table 2. The list of cryptocurrencies traded in Tokenomy. Data taken from Tokenomy.com on 5 November 2018.*

| No | Cryptocurrency | Code | Market | No | Cryptocurrency | Code | Market |
|----|----------------|------|--------|----|----------------|------|--------|
| 1 | Aeternity | AE | TEN | 16 | MidasProtocol | MAS | TEN, BTC, ETH |
| 2 | AppCoins | APPC | TEN, BTC, ETH | 17 | Mithril | MITH | TEN, BTC, ETH |
| 3 | Basic Attention Token | BAT | TEN, BTC, ETH | 18 | PundiX | NPXS | BTC, ETH |
| 4 | Bitcoin Cash | BCH | BTC | 19 | OmiseGO | OMG | BTC, ETH |
| 5 | Bread | BRD | BTC, ETH | 20 | Ontology | ONT | TEN, BTC |
| 6 | Bitcoin | BTC | USDT | 21 | Raiden Network Token | RDN | TEN, BTC, ETH |
| 7 | CyberMiles | CMT | TEN | 22 | Six | SIX | TEN, BTC |
| 8 | Daex | DAX | ETH | 23 | Status | SNT | TEN, BTC, ETH |
| 9 | Aelf | ELF | TEN | 24 | Storiqa | STQ | TEN, BTC, ETH |
| 10 | Ethereum Classic | ETC | BTC | 25 | Tokenomy | TEN | BTC, ETH, USDT |
| 11 | Ethereum | ETH | BTC, USDT | 26 | Tron | TRX | TEN, BTC, ETH |
| 12 | Golem | GNT | TEN, BTC, ETH | 27 | Veritaseum | VERI | BTC, ETH |
| 13 | Gifto | GTO | TEN, BTC, ETH | 28 | Vexanium | VEX | TEN, BTC, ETH |
| 14 | Loopring | LRC | TEN, BTC, ETH | 29 | Zilliqa | ZIL | TEN |
| 15 | Litecoin | LTC | BTC | 30 | 0x | ZRX | TEN |